\documentclass[fleqn,usenatbib]{mnras}

\usepackage{newtxtext,newtxmath}
\usepackage[T1]{fontenc}
\DeclareRobustCommand{\VAN}[3]{#2}
\let\VANthebibliography\thebibliography
\def\thebibliography{\DeclareRobustCommand{\VAN}[3]{##3}\VANthebibliography}

\usepackage{graphicx}	
\usepackage{amsmath}	
\usepackage{amssymb}	

\usepackage{times}

\def\gtorder{\mathrel{\raise.3ex\hbox{$>$}\mkern-14mu
             \lower0.6ex\hbox{$\sim$}}}
\def\ltorder{\mathrel{\raise.3ex\hbox{$<$}\mkern-14mu
             \lower0.6ex\hbox{$\sim$}}}

\title[Variable star candidates]{A catalog of over ten million variable source candidates in ZTF data release 1}

\author[Ofek et al.]{Eran O. Ofek\thanks{e-mail: eran.ofek@weizmann.ac.il}$^{1}$, Maayane Soumagnac$^{1,2}$, Guy Nir$^{1}$, Avishay Gal-Yam$^{1}$, \newauthor 
Peter Nugent$^{2}$, Frank Masci$^{3}$, Shrinivas R. Kulkarni$^{4}$\\
$^{1}$ Department of Particle Physics and Astrophysics, Weizmann Institute of Science, 76100 Rehovot, Israel\\
$^{2}$ Computational Cosmology Center, Lawrence Berkeley National Laboratory, 1 Cyclotron Road, Berkeley, CA 94720, USA\\
$^{3}$ Cahill Center for Astronomy and Astrophysics, California Institute of Technology, Pasadena, CA 91125, USA}

\date{Accepted ?
      Received ?
      in original form ?}

\begin{document}

\maketitle

\begin{abstract}

Variable sources probe a wide range of astrophysical phenomena.
We present a catalog of over ten million variable source candidates found in Data Release 1 (DR1) of the
Zwicky Transient Facility (ZTF).
We perform a periodicity search up to a frequency of 160\,day$^{-1}$,
and we classify the light curves into erratic and smooth variables.
We also present variability indicators and the results of a periodicity search,
up to a frequency of 5\,day$^{-1}$, for
about 1\,billion sources in the ZTF-DR1 light curve database.
We present several new short-period ($<90$\,min) candidates,
and about 60 new dwarf nova candidates, including two candidate eclipsing systems.
Both the 10 million variables catalog and $\sim1$ billion source catalog are available online
in {\tt catsHTM} format.

\end{abstract}


\section{Introduction}
\label{sec:intro}

Source variability allows us to probe a wide variety of astrophysical phenomena, from binary stars and exoplanets to stellar pulsations, cataclysmic and catastrophic events, and mass accretion by massive black holes.
In addition, stellar variability is of interest as it contaminates searches for transients.
Therefore, comprehensive lists of variable stars are of great importance.

Several productive variable star searches 
have been carried out. The Catalina Real Time Survey (CRTS)
published over 60,000 periodic variables
(e.g., \citealt{Drake+2014_CRTS_PeriodicVariables},
\citealt{Drake+2014_CRTS_CataclysmicVariables},
\citealt{Drake+2014_CRTS_UltraShortPeriodicVariables},
\citealt{Drake+2017_CRTS_SouthernPeriodicVariables}).
Other searches include
the OGLE survey (e.g., \citealt{Wozniak+2002_OGLE-II_BuldgeVariables}),
All Sky Automated Survey (e.g., \citealt{Pojmanski1997_ASAS_Description},
\citealt{Pojmanski2000_ASAS_Variables}),
Vista V\'{i}a L\'{a}ctea survey (VVV; e.g., 
\citealt{Minniti+2010_VVV_Survey}),
SDSS stripe 82 (e.g., \citealt{Sesar+2007_SDSS82_Variability}),
and more.
Recently, \cite{Chen+2020_ZTFDR2_PeriodicVariables} use the ZTF-DR2 to search periodic variables up to frequency of 40\,day$^{-1}$.
They identified about 781,000 periodic variable stars and classified them into several classes.

Here we present a search for variable star candidates in the Zwicky Transient Facility (ZTF; \citealt{Bellm+2019_ZTF_Overview}) Data Release 1 (DR1).
Variable stars are selected based on their large photometric flux root mean square (rms), compared to other stars in the field, or on some indication of periodicity in their light curves.
We provide the ZTF/DR1 photometric light curve catalog of about $1.6$ billion light curves,
of about a billion unique sources, in {\tt catsHTM} format (\citealt{Soumagnac+Ofek2018_catsHTM}).
For each source, we also list some variability attributes, calculate its periodogram up to a frequency of $5$\,day$^{-1}$, and provide the periodogram highest peak significance and frequency.
Next, we present a catalog of about 10.7 million variable star candidates.
For each variable candidate we provide a list of variability indicators and a periodicity search up to a frequency of $160$\,day$^{-1}$.

The structure of this paper is as follows.
In \S\ref{sec:catsHTM}, we describe the conversion of the ZTF light curves to binary HDF5 files,
and the {\tt catsHTM} catalog of sources and their variability properties.
In \S\ref{sec:var}, we describe the selection of variable star candidates,
while in \S\ref{sec:disc} we present some selected results.
We summarize in \S\ref{sec:sum}.

\section{ZTF-DR1 lightcurves in {\tt catsHTM} format}
\label{sec:catsHTM}

ZTF (\citealt{Bellm+2019_ZTF_Overview}; \citealt{Graham+2019_ZTF_objectives})
utilizes the 48-inch Schmidt telescope on Mount Palomar, equipped
with a 47\,deg$^{2}$ camera.
The ZTF data processing is described in \cite{Masci+2019_ZTF_Pipeline}.
The ZTF pipeline produces {\tt SExtractor} (\citealt{Bertin+1996_SExtractor})
and {\tt DAOPHOT} (\citealt{Stetson1987_DAOPHOT}) catalogs of sources in each image.
Here we use only the light curves based on the {\tt DAOPHOT} pipeline.
The ZTF photometry is calibrated against the Pan-STARRS1 catalog (\citealt{Chambers+2016_PS1_Surveys}).
A color term is fitted, but the magnitudes are calculated
assuming the sources have a color of $g-r=0$\,mag.

As part of its DR1, the ZTF collaboration released ascii files containing $1,681,215,104$ light curves of about $10^{9}$ non-unique sources
detected in the $g$ and $r$ bands.
The sources are non-unique, as they may be detected on several bands, and
in several ZTF fields.
Each ascii file contains all the $g$- and $r$-bands light curves of all the stars in one ZTF field (see \citealt{Bellm+2019_ZTF_Overview}, \citealt{Bellm+2019_ZTF_Scheduler}).
These files are available online\footnote{https://irsa.ipac.caltech.edu/data/ZTF/lc\_dr1/}.
We first converted these files into fast-access binary files in HDF5 format\footnote{https://www.hdfgroup.org/solutions/hdf5/},
and have made this HDF5 catalog public\footnote{https://euler1.weizmann.ac.il/catsHTM/}.
Python and MATLAB programs to access this catalog are available as part of the
{\tt catsHTM} toolbox.
For each light curve, we also provide some basic variability indicators and list the power and frequency of the highest peak in the classical periodogram (e.g., \citealt{Deeming1975_periodogram}) as calculated up to a frequency of 5\,day$^{-1}$.
The HDF5 file names are {\tt ztfLCDR1\_<field>.hdf5},
where {\tt <field>} is a zero padded six digit ZTF field number.
Each file contains two datasets\footnote{In the {\tt HDF5} terminology
a dataset can be regarded as a table of data.}.
The first dataset, named {\tt /AllLC}, is a matrix of all the light curves of all the stars in the field.
The light curves are stored one after another.
The matrix columns are listed in Table~\ref{tab:LCSD1} (see also the ZTF-DR1 online documentation\footnote{https://www.ztf.caltech.edu/page/dr1}).
The second dataset, named {\tt /IndAllLC}, is a matrix in which each line corresponds to a light curve of one source/filter in the first dataset.
For each source/filter, we store some basic information and also calculate some light curve statistics, and the frequency and power of the highest peak
in its periodogram.
The matrix columns are listed in Table~\ref{tab:LCSD2}.
Among the columns are the indices of the first and last line of the source's light curve in the first dataset (i.e., {\tt /AllLC}).

\begin{table}
    \centering
    \caption{Columns in the ZTF/DR1 light curve database.}
    \label{tab:LCSD1}
    
    \begin{tabular}{ll}
    \hline
    Column & Description \\
    \hline
    \hline
    HMJD       & Heliocentric modified Julian Day \\ 
    Mag        & ZTF Magnitude \\
    MagErr     & Magnitude error \\
    ColorCoef  & Color coeficent \\
    Flags      & 16bit flag \\
    \hline
    \end{tabular}
\end{table}

\begin{table*}
\begin{tabular}{ll}
\hline
Column & Description \\
\hline
\hline
RA        & J2000.0 Right Ascension [radians]\\
Dec       & J2000.0 Declination [radians]\\
I1        & First line number of source light curve in the light curve HDF5 file\\ 
I2        & Last line number\\
Nep       & Number of epochs \\
ID        & Source ID\\
FilterID  & Filter ID (1- $g$; 2- $r$)\\
Field     & Field ID\\
RcID      & CCD/quadrant ID\\
MeanMag   & Mean magnitude of source over all epochs\\
StdMag    & StD of source magnitude\\
RStdMag   & Robust StD of source magnitude\\
MaxMag    & Maximum magnitude minus mean magnitude\\
MinMag    & Mean magnitude minus minimum magnitude\\
Chi2      & $\chi^{2}$, where Nep is the number of degrees of freedom\\
MaxPower  & The power (in units of sigma) of the highest peak in the periodogram\\
FreqMaxPower & Frequency of maximal periodogram peak\\
\hline
\end{tabular}
\caption{Columns in the {\tt /IndAllLC} dataset, in the HDF files of all the ZTF-DR1 light curves, and in the {\tt catsHTM} catalog of all ZTF-DR1 sources. $\chi^{2}$ is the total chi-squared, calculated using the ZTF errors, where the number of degrees of freedom is $Nep$ minus one. The classical periodogram was calculated in frequency range of 0 to 5\,day$^{-1}$, with frequency steps of $1/600$\,day$^{-1}$.}
\label{tab:LCSD2}
\end{table*}

\begin{figure}
\includegraphics[width=84mm]{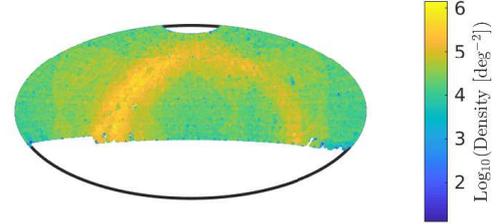}
\caption{The sky surface density of sources in the ZTF-DR1 light curve catalog, in Aitoff projection.
The source density is not uniform within a ZTF field/CCD quadrant, where the number of sources is slightly
lower toward the edges of the CCD quadrants.}
\label{fig:StarDensityZTFDR1}
\end{figure}

Next, we generated a {\tt catsHTM}-format (\citealt{Soumagnac+Ofek2018_catsHTM}) catalog of all the sources/filters with their properties (as listed in Table~\ref{tab:LCSD2}).
For each source/filter, we also stored the exact location of its light curve in the HDF5 light curves file.
This combination of {\tt catsHTM} and {\tt HDF5} files provides rapid search capabilities by coordinates and fast access to the entire light curve data.
Figure~\ref{fig:StarDensityZTFDR1} shows the sky source density of this catalog.

The reformatted catalogs are available from the {\tt catsHTM} website\footnote{https://euler1.weizmann.ac.il/catsHTM/}.
{\tt catsHTM} access tools are available in MATLAB\footnote{https://webhome.weizmann.ac.il/home/eofek/matlab/doc/catsHTM.html}
and Python\footnote{https://github.com/maayane/catsHTM}.

\section{Variable star selection and properties}
\label{sec:var}

Here we describe the variable star candidates selection, and their variability properties.

\subsection{Candidate Selection}

\begin{figure}
\includegraphics[width=84mm]{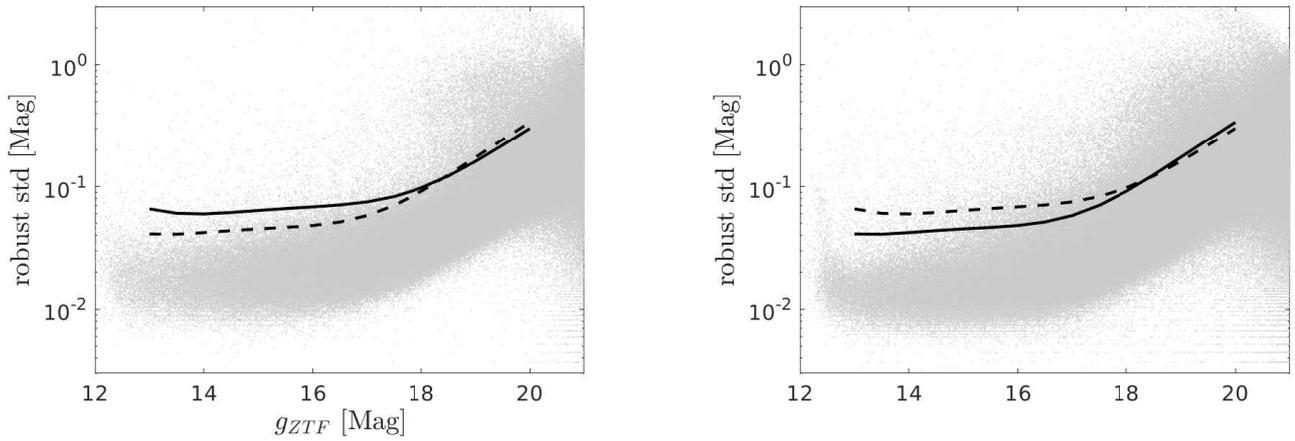}
\caption{The $g$-band robust rms of magnitude measurements of a source vs. source mean magnitude for about $3.9\times10^6$ random sources over the entire ZTF footprints. The solid line is the 4th degree
polynomial fit plus six times the rms, calculated in $0.1$ magnitude bins. The dashed line is like the solid line but for the $r$-band data. The lines-data is given in Table~\ref{tab:threshold4poly}.}
\label{fig:Mag_rstd_gZTF}
\end{figure}

\begin{figure}
\includegraphics[width=84mm]{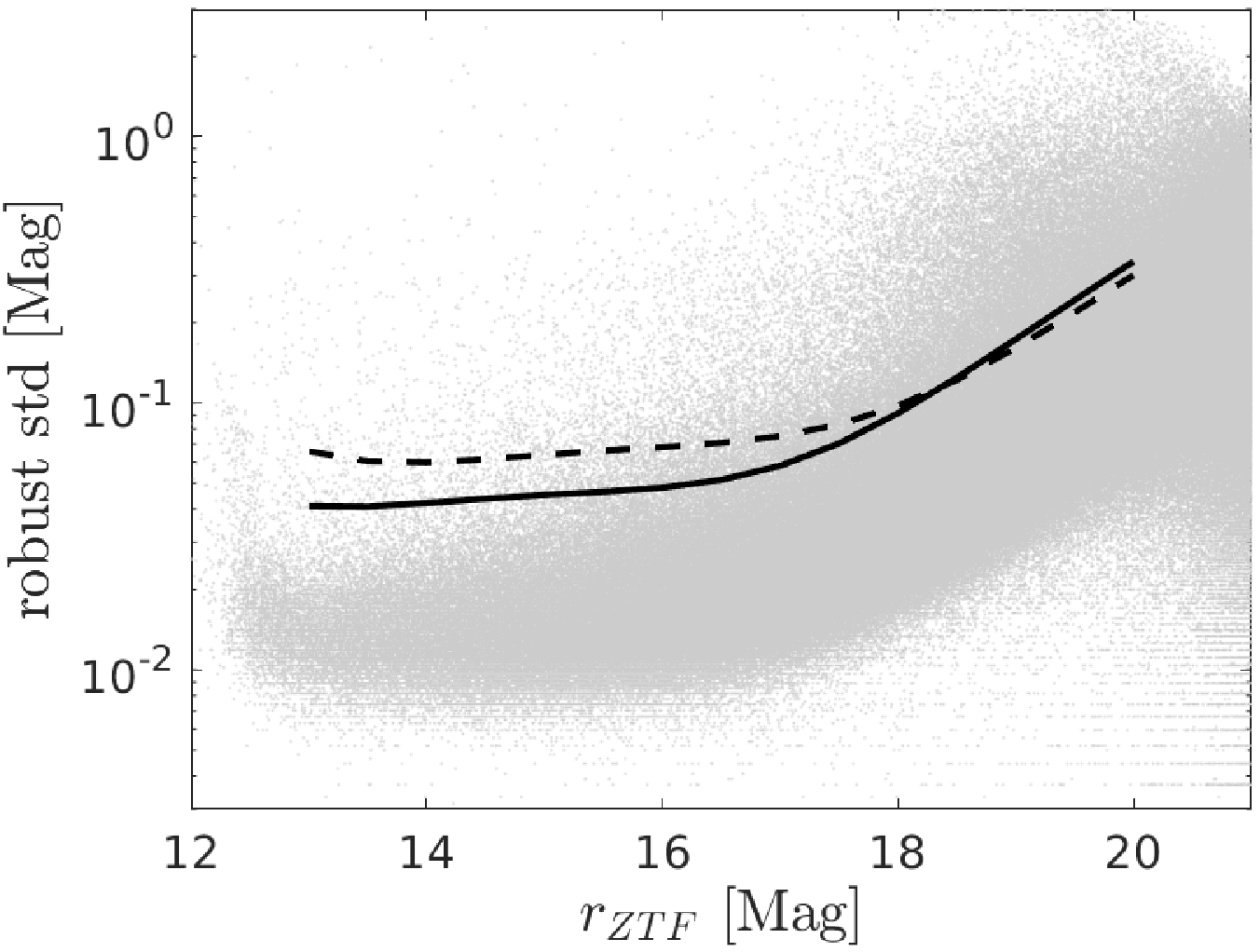}
\caption{Like Figure~\ref{fig:Mag_rstd_gZTF} but for the $r$-band. The dashed line is for the $g$-band.}
\label{fig:Mag_rstd_rZTF}
\end{figure}

We first estimated the sky-averaged magnitudes rms of sources as a function of magnitude.
In order to estimate this, we randomly selected 40,000 pointings across the sky, and for each pointing, we retrieved all the sources within 100\,arcsec from its position.
In total, about $3.9\times10^{6}$ sources were selected within the ZTF footprints.
Figures~\ref{fig:Mag_rstd_gZTF} and \ref{fig:Mag_rstd_rZTF}
show, in gray points, the robust\footnote{Calculated from the central 50\% percentile of data and normalized to 1-$\sigma$ assuming a Gaussian distribution.} standard deviation (StD) of the light curves of these stars as a function of their mean magnitude
in the $g$- and $r$-band, respectively.
For the region between magnitude 13 and 20, in magnitude bins of 0.1\,mag, we calculated the median
and the robust StD of these points.
The solid lines show the 4-th degree polynomial fitted to
the median-plus-six-times-the-robust-StD for the respective band,
while the dashed line is the same but for the other band.
The approximate 4-th degree polynomials are given in Table~\ref{tab:threshold4poly}.
We selected variable-star candidates based on their
high value of robust StD compared to other stars
with similar magnitudes.
A caveat of this approach is that we assumed the StD vs. magnitude
does~not depend on sky position.
In practice, this is likely not accurate.


\begin{table}
\begin{tabular}{lcc}
\hline
Mag & $g$-band threshold & $r$-band threshold \\
(mag) & (mag) & (mag)\\
\hline
\hline
           13   &  0.0658 &  0.0411 \\
         13.5   &  0.0605 &  0.0409 \\
           14   &  0.0598 &  0.0421 \\
         14.5   &  0.0615 &  0.0437 \\
           15   &  0.0639 &  0.0452 \\
         15.5   &  0.0662 &  0.0464 \\
           16   &  0.0683 &  0.0481 \\
         16.5   &  0.0708 &  0.0514 \\
           17   &  0.0751 &  0.0581 \\
         17.5   &  0.0831 &  0.0705 \\
           18   &  0.0976 &  0.0915 \\
         18.5   &  0.1222 &  0.1245 \\
           19   &  0.1609 &  0.1736 \\
         19.5   &  0.2189 &  0.2435 \\
           20   &  0.3016 &  0.3393 \\
\hline
\end{tabular}
\caption{The thresholds used for the selection of variable candidates. These are the six sigma lines in the $g$- and $r$-band filters shown in Figures~\ref{fig:Mag_rstd_gZTF} and \ref{fig:Mag_rstd_rZTF}.}
\label{tab:threshold4poly}
\end{table}

Next, we selected variable-candidate light curves, based on two criteria:
(i) Variable stars whose robust StD is larger than a band-dependent, magnitude-dependent threshold,
and whose number of data points (in the selected field/band) is larger than 24.
The magnitude-dependent threshold was selected to be the black-solid lines in 
Figures~\ref{fig:Mag_rstd_gZTF} and \ref{fig:Mag_rstd_rZTF} (see Table~\ref{tab:threshold4poly});
or (ii) Stars for which the highest peak in the classical periodogram calculated up to 5\,day$^{-1}$ is above 12 (normalized to the noise)
and the number of data points is larger than 24.
These two criteria left us with 12,761,565 (non-unique sources) candidate light curves.

\begin{figure}
\includegraphics[width=84mm]{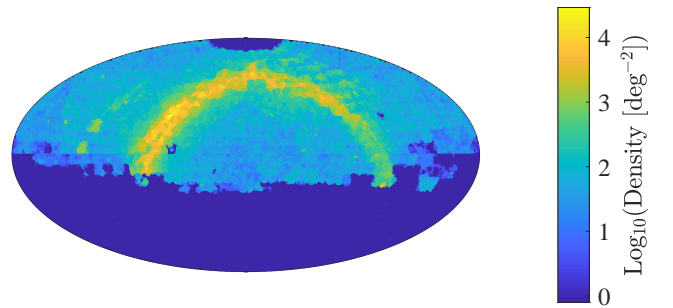}
\caption{The sky surface density of variable candidates in the ZTF-DR1 light curve catalog, in Aitoff projection.
There are some regions outside the Galactic plane with anomalous variable candidates surface density. This is likely due to some issues with the catalog (e.g., underestimation of the noise in some specific fields).}
\label{fig:VarDensityZTFDR1}
\end{figure}

Next, we merged the sources with light curves into unique sources, if the angular distance between the sources
was smaller than 1.5\,arcsec.
This merging process left us with
10,790,224 variable star candidates.
Out of the $\approx10.8$ million stars,
9,805,990 have one or more counterparts within $1.5''$ in the GAIA-DR2 catalog,
while 8,530,535 have exactly one counterpart within $1.5''$ in the GAIA-DR2 catalog.
The stars without a counterpart are typically bad detections (e.g., ghosts, bad pixels) or transient sources.
Figure~\ref{fig:VarDensityZTFDR1} shows the surface density of variable candidate on the celestial sphere.
There are some regions outside the Galactic plane with anomalous variable-candidates surface density. This is likely due to some issues with the catalog (e.g., underestimation of the noise in some specific fields).

For these $\approx10.8$ million stars, we calculated several additional properties, discussed in \S\ref{sec:prop}.
This catalog of variable star candidates along with their variability properties
is available in {\tt catsHTM} format\footnote{https://euler1.weizmann.ac.il/catsHTM/} (\citealt{Soumagnac+Ofek2018_catsHTM}).

\subsection{Properties of variable star candidates}
\label{sec:prop}

For each variable star candidate, we calculated additional variability properties.
The full list of a attributes in our variable-star candidates catalog is presented in Table~\ref{tab:VarProp}.
Here we discuss selected properties
and present their distributions.
\begin{table*}
\begin{tabular}{llll}
\hline
Column & Property & Units & Explanation \\
\hline
\hline
 1 &                   RA & deg & J2000.0 Right Ascension in the ZTF catalog (Epoch $\approx$J2018)  \\ 
 2 &                  Dec & deg & J2000.0 Declination in the ZTF catalog (Epoch $\approx$J2018)  \\ 
 3 &                 Napp &     & Number of matched sources in ZTF-DR1   \\ 
 4 &               SelBit &     & Flag indicating how the source was selected (e.g., STD, periodogram)  \\ 
 5 &              NcatDR1 &     & Number of ZTF-DR1 LCs within $1.5''$  \\ 
 6 &                 Nobs &     & Number of photometric observations in selected LCs  \\ 
 7 &              Ntotobs &     & Total number of photometric observations (all fields/filters)  \\ 
 8 &               Filter &     & Selected filter (1-$g$; 2-$r$).  \\ 
 \hline
 9 &              MeanMag & mag & Mean magnitude in selected LC.  \\ 
10 &               MedMag & mag & Median magnitude  \\ 
11 &              MeanErr & mag & Mean error  \\ 
12 &               MedErr & mag & Median error  \\ 
13 &               StdMag & mag & StD of magnitudes  \\ 
14 &               MinMag & mag & Minimum magnitude  \\ 
15 &               MaxMag & mag & Maximum magnitude  \\ 
16 &              RstdMag & mag & Robust StD of magnitudes  \\ 
17 &               RangeT & day & Time range  \\ 
18 &                 StdT & day & StD of time  \\ 
19 &                 Chi2 &     & $\chi^{2}$ of light curve, where Nobs$-1$ is the number of d.o.f.  \\ 
20 &    Corr\_MagColorCoef &     & Correlation coefficient between mag and color term \\ 
\hline
21 &            Mode1frac &     &  Fraction of measurements in the 1st most populated magnitude bin\\ 
22 &             Mode1mag & mag & Magnitude of the 1st most populated magnitude bin\\ 
23 &            Mode2frac &     & Fraction of measurements in the 2nd most populated magnitude bin  \\ 
24 &             Mode2mag & mag & Magnitude of the 2nd most populated magnitude bin\\ 
25 &            Mode3frac &     & Fraction of measurements in the 3rd most populated magnitude bin\\ 
26 &             Mode3mag & mag & Magnitude of the 3rd most populated magnitude bin\\ 
27 &            StdPoly10 & mag & StD of residuals from fiting 10th degree polynomial\\ 
28 &          MaxPSpower1 &     & Power of 1st highest peak in periodogram \\ 
29 &           MaxPSfreq1 & day$^{-1}$ & Frequency of 1st highest peak in periodogram  \\ 
30 &          MaxPSpower2 &    & Power of 2nd highest peak in periodogram\\ 
31 &           MaxPSfreq2 & day$^{-1}$ &  Frequency of 2nd highest peak in periodogram\\ 
32 &          MaxPSpower3 &    & Power of 3rd highest peak in periodogram\\ 
33 &           MaxPSfreq3 & day$^{-1}$ &  Frequency of 3rd highest peak in periodogram\\ 
34 &          MaxPSpower4 &   & Power of 4th highest peak in periodogram\\ 
35 &           MaxPSfreq4 & day$^{-1}$ &  Frequency of 4th highest peak in periodogram\\ 
36 &          MaxPSpower5 &   & Power of 5th highest peak in periodogram\\ 
37 &           MaxPSfreq5 & day$^{-1}$ &  Frequency of 5th highest peak in periodogram\\ 
\hline
38 &                Ngaia &   & Number of counterparts in GAIA-DR2 within $1.5''$\\ 
39 &                 MagG & mag  & GAIA $G$-magnitude of nearest source\\ 
40 &                MagBp & mag  & GAIA $B_{\rm p}$-magnitude   \\ 
41 &                MagRp & mag  & GAIA $R_{\rm p}$-magnitude   \\ 
42 &                  Plx & mas  & GAIA-DR2 parallax   \\ 
43 &               PlxErr & mas  & GAIA-DR2 parallax error   \\ 
44 &                   PM & mas\,yr$^{-1}$  & GAIA-DR2 total proper motion  \\ 
45 &          ExcessNoise & mas  & GAIA-DR2 excess noise   \\ 
\hline
46 &               z\_SDSS &   & SDSS redshift \\ 
47 &            zErr\_SDSS &   & SDSS redshift error \\ 
48 &           class\_SDSS &   & SDSS-spectrum object class    \\ 
49 &        subClass\_SDSS &   & SDSS-spectrum object subclass    \\ 
\hline
50 &       objType\_LAMOST &   & LAMOST-spectrum object type    \\ 
51 &         class\_LAMOST &   & LAMOST-spectrum object class     \\ 
52 &      subClass\_LAMOST &   & LAMOST-spectrum object subclass     \\ 
53 &             z\_LAMOST &   & LAMOST redshift    \\ 
54 &          zErr\_LAMOST &   & LAMOST redshift error \\ 
\hline
\end{tabular}
\caption{Columns available in the ZTF-DR1 variable candidates catalog.}
\label{tab:VarProp}
\end{table*}

{\bf The five highest peaks in the periodogram:}
We re-calculated the classical periodogram, this time up to frequency of 160\,day$^{-1}$.
We choose to calculate the classical periodogram, and not
the Lomb-Scargle periodogram (\citealt{Lomb1976_periodogram_LeastSquares}, \citealt{Scargle1982_periodogram}),
due to speed considerations.
We kept in the catalog the height and frequency of the five highest peaks in the periodogram.

\begin{figure}
\includegraphics[width=84mm]{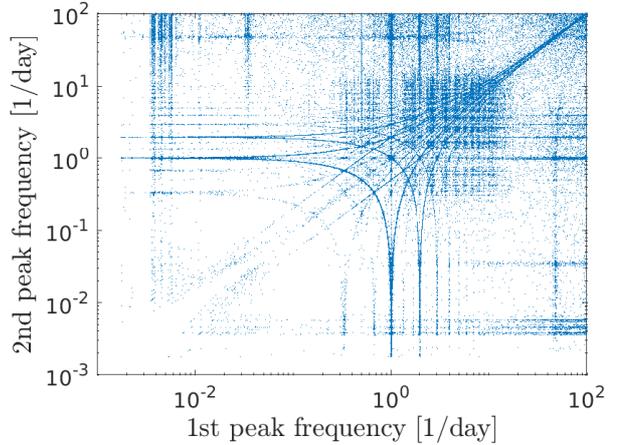}
\caption{The first highest peak frequency vs. second highest peak frequency in the periodogram,
for sources in which the highest periodogram peak is above 12.
As discussed in the text, the over abundance in specific frequencies is due to the aliasing of a true period with the window function of the data.}
\label{fig:Var_PS_f1_f2}
\end{figure}

In Figure~\ref{fig:Var_PS_f1_f2}, we show the first highest-peak frequency vs. second highest-peak frequency in the periodogram
for 5,400,501 sources for which the highest periodogram peak is above 12.
The vertical, and horizontal lines that are over abundant in sources are mainly due to
the window function of the data and aliases with the window function.
The lines near frequency of 0.004 to 0.006\,day$^{-1}$ corresponds to several months,
which is the typical yearly observation period of ZTF per source.
The line near the frequency of 0.03\,day$^{-1}$ corresponds to the lunar synodic period.
Next, we can see a series of lines at about 0.33, 0.5, 1, 2, ..., and up to 13\,day$^{-1}$.
These lines correspondsto the whole multiplicities of the sidereal day period (i.e., $0.997$\,day).
Finally, the thick line near 48\,day$^{-1}$ corresponds to the ZTF minimal mean cadence of about 30\,min.
The concentration of sources near the whole multipliers of the sidereal day period is mainly due to periodic variables
in which the true period (typically with long periods) generated a strong alias with the sidereal period.
In fact, in such cases, two peaks are generated, one below and one above the sidereal period,
and the frequencies of these two peaks are
$1.013\pm 1/P$\,day$^{-1}$, where $P$ is the true periodicity and $1.013$\,day$^{-1}$ is the sidereal day frequency.
This explains the bifurcation of many of the lines clearly seen in this diagram, and it allows us to estimate the
true periodicity of such sources.

\begin{figure}
\includegraphics[width=84mm]{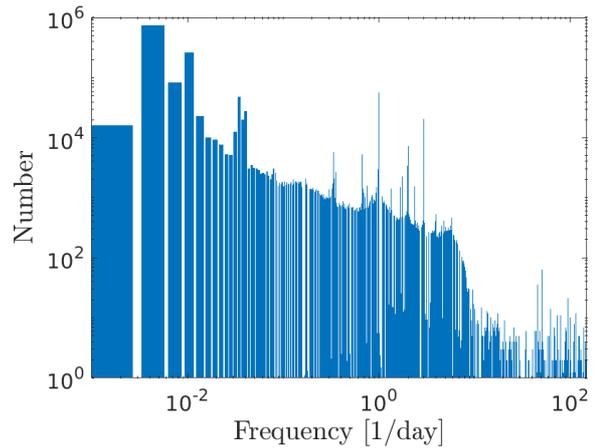}
\caption{The histogram of the highest peak frequency, for sources in which the power of the highest
peak is larger than 12.}
\label{fig:Var_PS_hist_f1}
\end{figure}

Figure~\ref{fig:Var_PS_hist_f1} presents the histogram of the highest peak frequency, for sources in which the power of the highest
peak is larger than 12.
The bins are per 0.003\,day$^{-1}$ frequency, but are plotted in logarithmic scale.
The peaks discussed in the context of Figure~\ref{fig:Var_PS_f1_f2} are clearly visible.
The jump in the abundance (and also the power, see Figure~\ref{fig:Var_PS_hist_f1}) of sources around frequency of 10\,day$^{-1}$
is likely due to loss of efficiency above this frequency (due to the ZTF cadence), and the fact
we are using the classical periodogram, which, unlike the Lomb-Scargle periodogram (\citealt{Lomb1976_periodogram_LeastSquares}), does~not normalize the power to the frequency response of the
periodogram to a sine wave at that frequency.

\begin{figure}
\includegraphics[width=84mm]{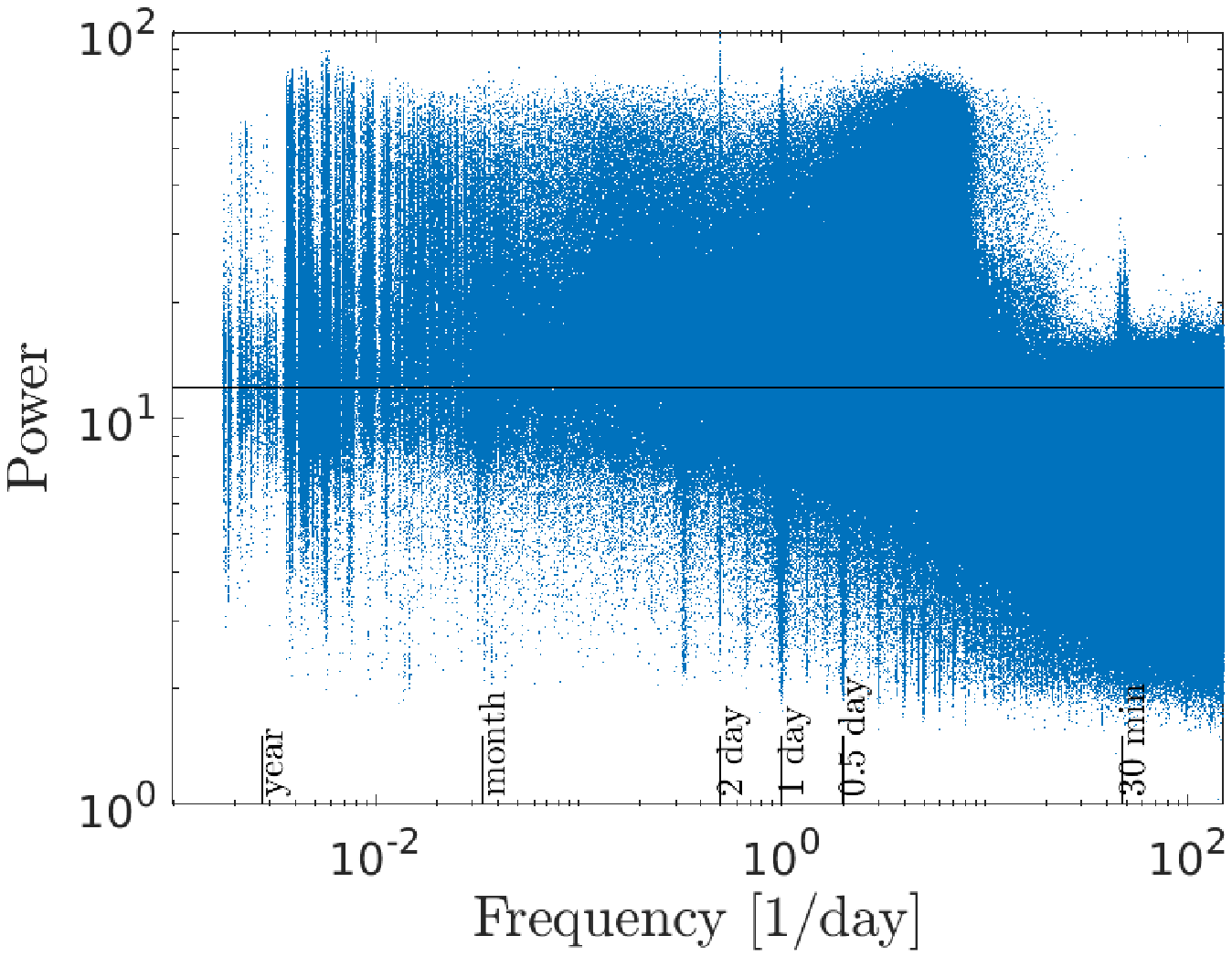}
\caption{The frequency vs. power of the highest peak in the periodogram for the sources in the variable candidates catalog. The power is normalized by the variance, such that it roughly corresponds to the significance of the peak in units of the noise. The solid line shows the power equal 12 level.}
\label{fig:Var_PS_f1_p1}
\end{figure}

Figure~\ref{fig:Var_PS_f1_p1} shows the periodogram highest peak vs. power. The solid black line marks
our periodicity threshold (12), while prominent frequencies are marked at the bottom.

{\bf Standard deviation after a high-order polynomial fit:}
We subtracted from the time of each light curve the mean time and divided it by the time range. Each time-normalized light curve was fitted with a 10th degree polynomial. Next, we calculated the standard deviation of the polynomial-subtracted light curve (i.e., {\it StdPoly10}).
The motivation for obtaining this property is that variable stars with smooth and slowly evolving
light curves (e.g., Mira stars) can be easily identified by their low StdPoly10, while eruptive variables (e.g., Dwarf novae) tend to have a high standard deviation even
after the removal of such a polynomial.
The response of this estimator to eruptive variables is not uniform.
The reason is that different sources were observed with a different time range and sampling.

\begin{figure}
\includegraphics[width=84mm]{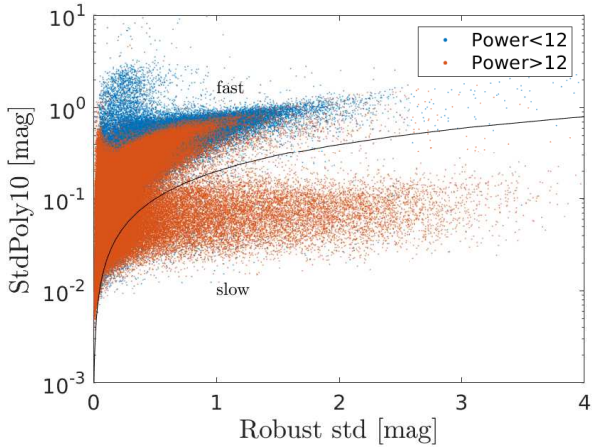}
\caption{The robust StD vs. the {\it StDPoly10} (i.e., StD after fitting and subtracting a 10th degree polynomial from the light curve).
Blue points correspond to sources whose highest peak in the periodogram is below 12,
while red points denote sources whose highest peak in the periodogram is above 12.}
\label{fig:Var_rstd_poly10std}
\end{figure}

Figure~\ref{fig:Var_rstd_poly10std} shows the robust StD vs. the StD after fitting and removing the 10th degree polynomial from the light curve.
The solid black line corresponds to $y=0.2x$, where $x$ and $y$ are the plot axes.
Blue points correspond to sources in which the highest peak in the periodogram is below 12,
while red points denote sources in which the highest peak in the periodogram is above 12.
This plot distinguishes between fast variables (mainly above the black line),
and slow variables (mainly below the black line).
This plot shows that slow variables are mostly periodic. Furthermore, periodic variables that are classified as fast tend to have low amplitudes compared to the population of fast variables.

{\bf Peaks in the magnitude histogram:}
For each light curve, we calculated a histogram of the magnitude measurements with $0.25$\,mag bins.
We selected the three bins with the highest number of magnitude measurements.
For each one of these three bins, we stored the magnitude at the middle of the bin (i.e., {\tt mode1mag}, {\tt mode2mag}, {\tt mode3mag})
and the fraction of measurements in the bin compared to the total number of measurements
(i.e., {\tt mode1frac}, {\tt mode2frac}, {\tt mode3frac}).
These indicators maybe able to assist in identifying variable stars that mainly move from
one state to another; examples include eclipsing binaries and dwarf novae.
It can also be used to identify stars that are likely flagged as variable candidates due to ouliers
or due to the mistaken matching of multiple sources as a single star.

\section{selected results}
\label{sec:disc}

Here we present some selected results related to variable stars.
In \S\ref{sec:shortperiod} and \S\ref{sec:dwarfnovae}, we present a preliminary search for
short-period variables, and dwarf novae, respectively.

\subsection{short period variability}
\label{sec:shortperiod}

Given the typical minimum cadence of ZTF (about 30\,min),
our period search is not very effective above frequencies of $12$\,day$^{-1}$
(i.e., Nyquist frequency for 30\,min sampling).
Since the ZTF cadence is not strict, but has some randomness,
it is possible to identify periods shorter than about 1\,hr.
However, any such search for short-period variability will have low
efficiency and a high false-alarm rate.

\begin{figure*}
\includegraphics[width=160mm]{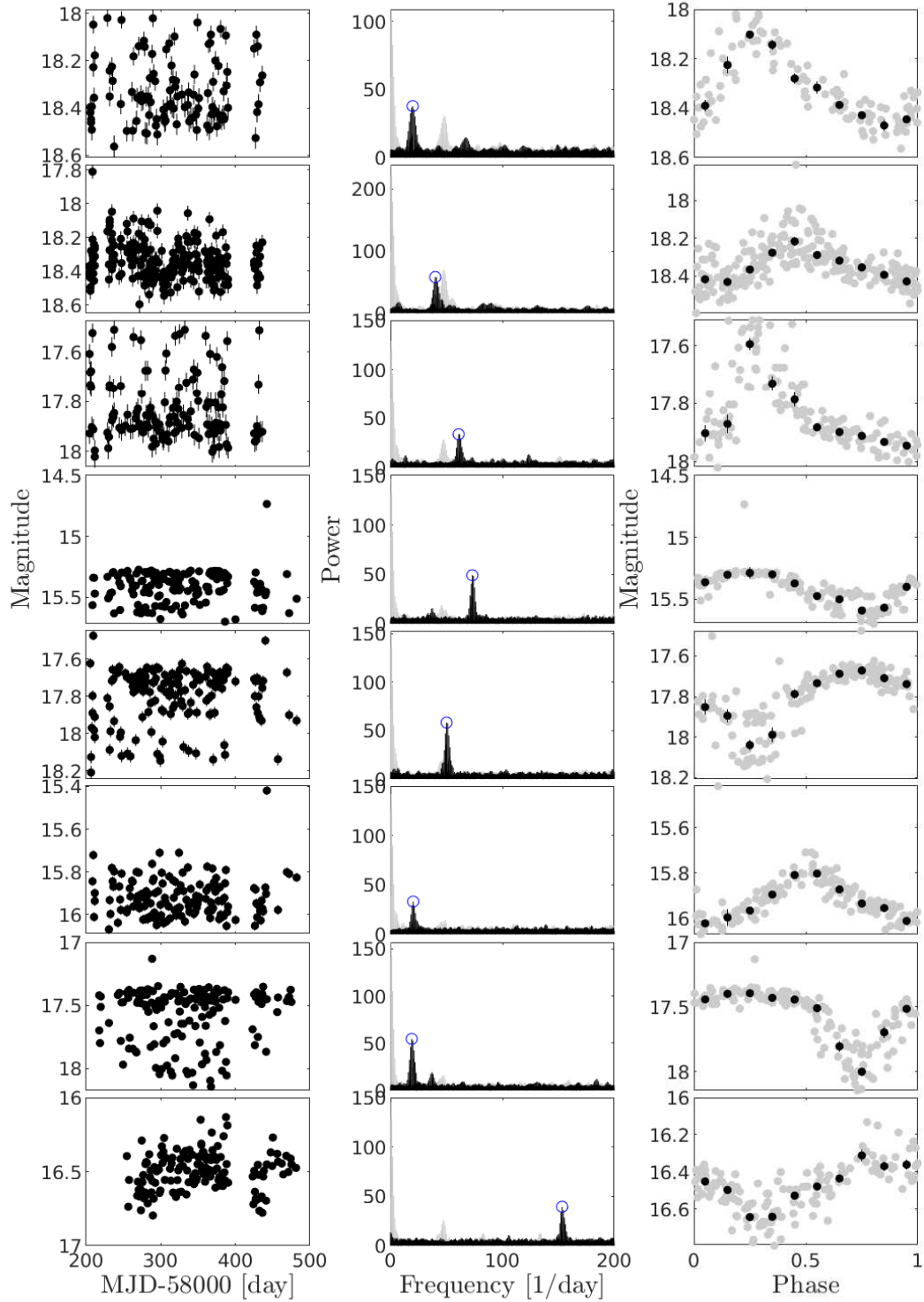}
\caption{The light curves of eight short-period ($<90$\,min) variable candidates -- row per source. The left column shows the light curves. The middle column presents the power spectra (black), the window functions (gray), and the strongest power-spectrum peaks (blue circles). The right column shows the light curves folded into the period (strongest peak in the periodogram). Gray points are the measurements while black dots are median magnitudes in bins of 0.1 of the period. Candidate coordinates and periods are listed in Table~\ref{tab:shortperiod} (in the same order).}
\label{fig:ZTF_DR1_Var_ShortPeriod}
\end{figure*}

In order to test the possibility of identifying short-period variable candidates, we selected sources
using the following criteria:
(i) The power of the highest peak in the periodogram $>25$.
(ii) The frequency of the first and second highest peaks in the periodogram are $>18$\,day$^{-1}$.
(iii) Robust StD $>0.02$\,mag.
(iv) StdPoly10 $>1.1$.
(v) Magnitude $G<18$.
This resulted in 63 candidates. We inspected all these candidates by eye and found eight promising short-period candidates.
The properties of these candidates are listed in Table~\ref{tab:shortperiod},
while their light curves, periodograms, and folded light curves,
are presented in Figure~\ref{fig:ZTF_DR1_Var_ShortPeriod}.

\begin{table*}
\begin{tabular}{rrllrl}
\hline
J2000.0 R.A. & J2000.0 Dec. & Filter & Med. Mag. & Period & Comments \\
(deg) & (deg) & () & (mag) & (min) & () \\
\hline
290.81069 &  3.89778 &$g$ &18.36&  72.44524 &\\
288.28656 & 12.08099 &$g$ &18.36&  36.01894 & \\
288.67014 & 19.64048 &$g$ &17.87&  23.64739 & RGB star\\
322.73628 & 44.34625 &$g$ &15.40&  19.66984 & MGAB-V249 - AM CVn \\
313.81656 & 46.85180 &$g$ &17.75&  28.73726 & nearby X-ray source\\
320.62207 & 57.32472 &$g$ &15.94&  70.86444 & \\
326.03118 & 58.30484 &$r$ &17.47&  75.75160 & \\
  5.74011 & 61.68544 &$r$ &16.49&   9.38529 & V1033 Cas (Nova)\\
\hline
\end{tabular}
\caption{List of selected short-period ($<90$\,min) variable candidates found in the ZTF variability catalog. Med. Mag. is the median magnitude in the ZTF band. Period accuracy is about a few miliseconds. The comments are based on information from SIMBAD (\citealt{Wenger+2000_SIMBAD}).}
\label{tab:shortperiod}
\end{table*}

\subsection{Dwarf novae candidates}
\label{sec:dwarfnovae}

\begin{figure*}
\includegraphics[width=160mm]{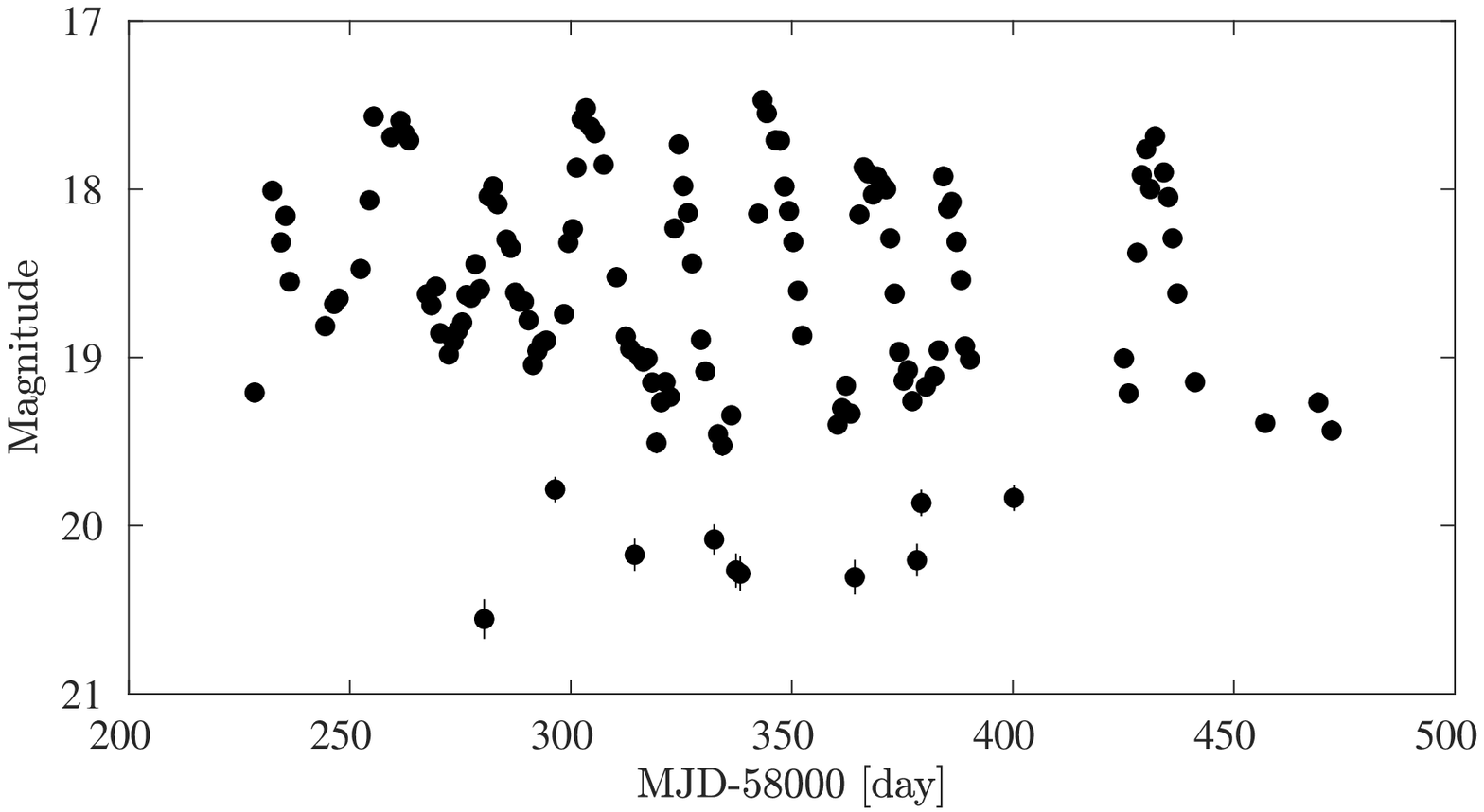}
\caption{The light curve of the best eclipsing dwarf nova candidate,
ZTF J$222959.198+521507.78$.
A possible period for the eclipses is 5.9186\,h. However, this is not a unique solution and further observations are required.}
\label{fig:DN_withEclipse}
\end{figure*}

We conducted a partial search for dwarf nova eruptions.
We used the following selection criteria:
(i) The power of the highest peak in the periodogram $>8$.
(ii) The frequency of the highest peak in the periodogram is between $0.01$ and $0.1$\,day$^{-1}$.
(iii) Robust StD $>0.2$\,mag.
(iv) StdPoly10 $>0.3$\,mag.
(v) Magnitude range $>2.5$\,mag.
This resulted in 452 candidates.
We inspected the candidates light curves by eye and selected 76 dwarf novae (DN) candidates.
The selection was based on the similarity of the light curves to other known classical dwarf novae.
About 60 of the candidates were previously unknown.
The candidates are listed in Table~\ref{tab:dn_cand}.
Two candidates show some evidence of eclipses.
The light curve of the best eclipsing dwarf nova candidate, ZTF J$222959.198+521507.78$, is shown in Figure~\ref{fig:DN_withEclipse}.

\begin{table*}
\begin{tabular}{rrcllll}
\hline
J2000.0 R.A. & J2000.0 Dec. & Filter & Min. Mag. & Max. Mag. & Med. Mag. & Comments \\
(deg) & (deg) & () & (mag) & (mag) & (mag) & () \\
\hline
313.82849 & $-16.44455$ & $r$ & 16.65 & 20.19 & 18.74 &   SDSS J205518.83-162640.4 \\ 
109.17199 & $ -6.94683$ & $g$ & 13.98 & 19.75 & 18.95 &   FQ Mon (Mira cand), but looks like a DN \\ 
 99.22749 & $  0.03812$ & $g$ & 12.37 & 17.29 & 16.65 &   CW Mon (DN) \\ 
280.32467 & $  4.15413$ & $r$ & 17.81 & 20.90 & 20.36 &   \\ 
 93.38401 & $  6.95214$ & $g$ & 17.39 & 21.46 & 18.37 &   \\ 
\hline
\end{tabular}
\caption{List of selected dwarf novae (DN) candidates found in the ZTF variability catalog. First five lines are shown. The full table is available in the electronic version of the paper.}
\label{tab:dn_cand}
\end{table*}

\section{Summary}
\label{sec:sum}

We present a catalog of over ten milion variable star candidates identified in the ZTF-DR1 light curves database.
We provide the entire ZTF-DR1 light curves database in {\tt HDF5} format and an index catalog in {\tt catsHTM}
format. We also generated a catalog of variable candidates in {\tt catsHTM} format.
We provide some variability indicators for the variable star candidates, including a division into
erratic (fast) and smooth (slow) variable sources.

We conducted a preliminary search for variable sources with a periodicity shorter than 90\,min.
We identified eight candidates, one of which is a known AM~CVn star.
We also searched for outbursting dwarf novae, and found about 60 new dwarf novae candidates.
This catalog may also be useful for transient searches where a catalog of stellar variability helps in removing false alarms.

\section*{ACKNOWLEDGMENTS}
E.O.O. is grateful for the support by
grants from the Israel Science Foundation, Minerva, Israeli Ministry of Technology and Science, the US-Israel Binational Science Foundation, Weizmann-UK, Weizmann-Yale,
and the Weizmann-Caltech grants.

\bibliographystyle{mnras}
\bibliography{main.bib}

\begin{thebibliography}{}
\makeatletter
\relax
\def\mn@urlcharsother{\let\do\@makeother \do\$\do\&\do\#\do\^\do\_\do\%\do\~}
\def\mn@doi{\begingroup\mn@urlcharsother \@ifnextchar [ {\mn@doi@}
  {\mn@doi@[]}}
\def\mn@doi@[#1]#2{\def\@tempa{#1}\ifx\@tempa\@empty \href
  {http://dx.doi.org/#2} {doi:#2}\else \href {http://dx.doi.org/#2} {#1}\fi
  \endgroup}
\def\mn@eprint#1#2{\mn@eprint@#1:#2::\@nil}
\def\mn@eprint@arXiv#1{\href {http://arxiv.org/abs/#1} {{\tt arXiv:#1}}}
\def\mn@eprint@dblp#1{\href {http://dblp.uni-trier.de/rec/bibtex/#1.xml}
  {dblp:#1}}
\def\mn@eprint@#1:#2:#3:#4\@nil{\def\@tempa {#1}\def\@tempb {#2}\def\@tempc
  {#3}\ifx \@tempc \@empty \let \@tempc \@tempb \let \@tempb \@tempa \fi \ifx
  \@tempb \@empty \def\@tempb {arXiv}\fi \@ifundefined
  {mn@eprint@\@tempb}{\@tempb:\@tempc}{\expandafter \expandafter \csname
  mn@eprint@\@tempb\endcsname \expandafter{\@tempc}}}

\bibitem[\protect\citeauthoryear{{Bellm} et~al.,}{{Bellm}
  et~al.}{2019a}]{Bellm+2019_ZTF_Overview}
{Bellm} E.~C.,  et~al., 2019a, \mn@doi [\pasp] {10.1088/1538-3873/aaecbe},
  \href {https://ui.adsabs.harvard.edu/abs/2019PASP..131a8002B} {131, 018002}

\bibitem[\protect\citeauthoryear{{Bellm} et~al.,}{{Bellm}
  et~al.}{2019b}]{Bellm+2019_ZTF_Scheduler}
{Bellm} E.~C.,  et~al., 2019b, \mn@doi [\pasp] {10.1088/1538-3873/ab0c2a},
  \href {https://ui.adsabs.harvard.edu/abs/2019PASP..131f8003B} {131, 068003}

\bibitem[\protect\citeauthoryear{{Bertin} \& {Arnouts}}{{Bertin} \&
  {Arnouts}}{1996}]{Bertin+1996_SExtractor}
{Bertin} E.,  {Arnouts} S.,  1996, \mn@doi [\aaps] {10.1051/aas:1996164}, \href
  {https://ui.adsabs.harvard.edu/abs/1996A&AS..117..393B} {117, 393}

\bibitem[\protect\citeauthoryear{{Chambers} et~al.,}{{Chambers}
  et~al.}{2016}]{Chambers+2016_PS1_Surveys}
{Chambers} K.~C.,  et~al., 2016, arXiv e-prints, \href
  {https://ui.adsabs.harvard.edu/abs/2016arXiv161205560C} {p. arXiv:1612.05560}

\bibitem[\protect\citeauthoryear{{Chen}, {Wang}, {Deng}, {de Grijs}, {Yang}  \&
  {Tian}}{{Chen} et~al.}{2020}]{Chen+2020_ZTFDR2_PeriodicVariables}
{Chen} X.,  {Wang} S.,  {Deng} L.,  {de Grijs} R.,  {Yang} M.,   {Tian} H.,
  2020, arXiv e-prints, \href
  {https://ui.adsabs.harvard.edu/abs/2020arXiv200508662C} {p. arXiv:2005.08662}

\bibitem[\protect\citeauthoryear{{Deeming}}{{Deeming}}{1975}]{Deeming1975_periodogram}
{Deeming} T.~J.,  1975, \mn@doi [\apss] {10.1007/BF00681947}, \href
  {https://ui.adsabs.harvard.edu/abs/1975Ap&SS..36..137D} {36, 137}

\bibitem[\protect\citeauthoryear{{Drake} et~al.,}{{Drake}
  et~al.}{2014a}]{Drake+2014_CRTS_PeriodicVariables}
{Drake} A.~J.,  et~al., 2014a, \mn@doi [\apjs] {10.1088/0067-0049/213/1/9},
  \href {https://ui.adsabs.harvard.edu/abs/2014ApJS..213....9D} {213, 9}

\bibitem[\protect\citeauthoryear{{Drake} et~al.,}{{Drake}
  et~al.}{2014b}]{Drake+2014_CRTS_CataclysmicVariables}
{Drake} A.~J.,  et~al., 2014b, \mn@doi [\mnras] {10.1093/mnras/stu639}, \href
  {https://ui.adsabs.harvard.edu/abs/2014MNRAS.441.1186D} {441, 1186}

\bibitem[\protect\citeauthoryear{{Drake} et~al.,}{{Drake}
  et~al.}{2014c}]{Drake+2014_CRTS_UltraShortPeriodicVariables}
{Drake} A.~J.,  et~al., 2014c, \mn@doi [\apj] {10.1088/0004-637X/790/2/157},
  \href {https://ui.adsabs.harvard.edu/abs/2014ApJ...790..157D} {790, 157}

\bibitem[\protect\citeauthoryear{{Drake} et~al.,}{{Drake}
  et~al.}{2017}]{Drake+2017_CRTS_SouthernPeriodicVariables}
{Drake} A.~J.,  et~al., 2017, \mn@doi [\mnras] {10.1093/mnras/stx1085}, \href
  {https://ui.adsabs.harvard.edu/abs/2017MNRAS.469.3688D} {469, 3688}

\bibitem[\protect\citeauthoryear{{Graham} et~al.,}{{Graham}
  et~al.}{2019}]{Graham+2019_ZTF_objectives}
{Graham} M.~J.,  et~al., 2019, \mn@doi [\pasp] {10.1088/1538-3873/ab006c},
  \href {https://ui.adsabs.harvard.edu/abs/2019PASP..131g8001G} {131, 078001}

\bibitem[\protect\citeauthoryear{{Lomb}}{{Lomb}}{1976}]{Lomb1976_periodogram_LeastSquares}
{Lomb} N.~R.,  1976, \mn@doi [\apss] {10.1007/BF00648343}, \href
  {https://ui.adsabs.harvard.edu/abs/1976Ap&SS..39..447L} {39, 447}

\bibitem[\protect\citeauthoryear{{Masci} et~al.,}{{Masci}
  et~al.}{2019}]{Masci+2019_ZTF_Pipeline}
{Masci} F.~J.,  et~al., 2019, \mn@doi [\pasp] {10.1088/1538-3873/aae8ac}, \href
  {https://ui.adsabs.harvard.edu/abs/2019PASP..131a8003M} {131, 018003}

\bibitem[\protect\citeauthoryear{{Minniti} et~al.,}{{Minniti}
  et~al.}{2010}]{Minniti+2010_VVV_Survey}
{Minniti} D.,  et~al., 2010, \mn@doi [\na] {10.1016/j.newast.2009.12.002},
  \href {https://ui.adsabs.harvard.edu/abs/2010NewA...15..433M} {15, 433}

\bibitem[\protect\citeauthoryear{{Pojmanski}}{{Pojmanski}}{1997}]{Pojmanski1997_ASAS_Description}
{Pojmanski} G.,  1997, \actaa, \href
  {https://ui.adsabs.harvard.edu/abs/1997AcA....47..467P} {47, 467}

\bibitem[\protect\citeauthoryear{{Pojmanski}}{{Pojmanski}}{2000}]{Pojmanski2000_ASAS_Variables}
{Pojmanski} G.,  2000, \actaa, \href
  {https://ui.adsabs.harvard.edu/abs/2000AcA....50..177P} {50, 177}

\bibitem[\protect\citeauthoryear{{Scargle}}{{Scargle}}{1982}]{Scargle1982_periodogram}
{Scargle} J.~D.,  1982, \mn@doi [\apj] {10.1086/160554}, \href
  {https://ui.adsabs.harvard.edu/abs/1982ApJ...263..835S} {263, 835}

\bibitem[\protect\citeauthoryear{{Sesar} et~al.,}{{Sesar}
  et~al.}{2007}]{Sesar+2007_SDSS82_Variability}
{Sesar} B.,  et~al., 2007, \mn@doi [\aj] {10.1086/521819}, \href
  {https://ui.adsabs.harvard.edu/abs/2007AJ....134.2236S} {134, 2236}

\bibitem[\protect\citeauthoryear{{Soumagnac} \& {Ofek}}{{Soumagnac} \&
  {Ofek}}{2018}]{Soumagnac+Ofek2018_catsHTM}
{Soumagnac} M.~T.,  {Ofek} E.~O.,  2018, \mn@doi [\pasp]
  {10.1088/1538-3873/aac410}, \href
  {https://ui.adsabs.harvard.edu/abs/2018PASP..130g5002S} {130, 075002}

\bibitem[\protect\citeauthoryear{{Stetson}}{{Stetson}}{1987}]{Stetson1987_DAOPHOT}
{Stetson} P.~B.,  1987, \mn@doi [\pasp] {10.1086/131977}, \href
  {https://ui.adsabs.harvard.edu/abs/1987PASP...99..191S} {99, 191}

\bibitem[\protect\citeauthoryear{{Wenger} et~al.,}{{Wenger}
  et~al.}{2000}]{Wenger+2000_SIMBAD}
{Wenger} M.,  et~al., 2000, \mn@doi [\aaps] {10.1051/aas:2000332}, \href
  {https://ui.adsabs.harvard.edu/abs/2000A&AS..143....9W} {143, 9}

\bibitem[\protect\citeauthoryear{{Wozniak}, {Udalski}, {Szymanski}, {Kubiak},
  {Pietrzynski}, {Soszynski}  \& {Zebrun}}{{Wozniak}
  et~al.}{2002}]{Wozniak+2002_OGLE-II_BuldgeVariables}
{Wozniak} P.~R.,  {Udalski} A.,  {Szymanski} M.,  {Kubiak} M.,  {Pietrzynski}
  G.,  {Soszynski} I.,   {Zebrun} K.,  2002, \actaa, \href
  {https://ui.adsabs.harvard.edu/abs/2002AcA....52..129W} {52, 129}

\makeatother
\end{thebibliography}



\end{document}